\documentclass[prd,preprintnumbers,superscriptaddress,nofootinbib,floatfix,twocolumn,notitlepage]{revtex4}
\usepackage[dvips]{graphicx}
\usepackage{dcolumn} 
\usepackage{bm}
\usepackage{epsfig,amsmath,amssymb,verbatim,mathrsfs,array,layout,textcomp,amssymb,latexsym,slashed}
\usepackage{color}
\usepackage[colorlinks=true,citecolor=blue,urlcolor=blue,linktocpage=true,
linkcolor=blue]{hyperref}
\usepackage[utf8]{inputenc}
\usepackage{multirow}

\input epsf.tex
\newcommand{\beq}{\begin{eqnarray}}
\newcommand{\eeq}{\end{eqnarray}}

\def\ltap{\ \raise.3ex\hbox{$<$\kern-.75em\lower1ex\hbox{$\sim$}}\ }
\def\gtap{\ \raise.3ex\hbox{$>$\kern-.75em\lower1ex\hbox{$\sim$}}\ }

\def\be{\begin{equation}}
\def\ee{\end{equation}}
\def\bea{\begin{eqnarray}}
\def\eea{\end{eqnarray}}

\pdfoutput=1

\definecolor{newred}{rgb}{0.5,0.1,0}
\definecolor{darkgreen}{rgb}{0.0,0.7,0.2}
\definecolor{lightblue}{rgb}{0.0,0.5,1}

\newcommand{\delr}{\delta \langle r^2 \rangle}

\RequirePackage[normalem]{ulem} 
\RequirePackage{color}\definecolor{RED}{rgb}{1,0,0}\definecolor{BLUE}{rgb}{0,0,1} 

\begin{document} 

\title{Probing  new spin-independent  interactions through precision spectroscopy in atoms with few electrons}
\preprint{LAPTH-028/17, MIT-CTP/4929}

\author{C\'edric Delaunay}
\affiliation{Laboratoire d'Annecy-le-Vieux de Physique Th\'eorique LAPTh, CNRS -- USMB, BP 110 Annecy-le-Vieux, F-74941 Annecy, France}
\author{Claudia Frugiuele}
\affiliation{Department of Particle Physics and Astrophysics, Weizmann Institute of Science, Rehovot 7610001, Israel}
\author{Elina Fuchs}
\affiliation{Department of Particle Physics and Astrophysics, Weizmann Institute of Science, Rehovot 7610001, Israel}
\author{Yotam Soreq}
\affiliation{Center for Theoretical Physics, Massachusetts Institute of Technology, Cambridge, MA  02139, U.S.A.}

\begin{abstract}

The very high precision of current measurements and theory predictions of spectral lines in few-electron atoms  allows to efficiently probe the existence of exotic forces between electrons, neutrons and protons. We investigate the sensitivity to new spin-independent interactions in transition frequencies (and their isotopic shifts) of hydrogen, helium and some helium-like ions.
We find that present data probe new regions of the force-carrier couplings to electrons and neutrons  around the MeV mass range.
We also find that, below few keV, the sensitivity to the electron coupling in precision spectroscopy of helium and positronium is comparable to that of the anomalous magnetic moment of the electron. 
Finally, we interpret our results in the dark-photon model where a new gauge boson is kinetically mixed with the photon. There, we show that helium transitions, combined with the anomalous magnetic moment of the electron,
provide the strongest indirect bound from laboratory experiments above $100\,$keV.
\end{abstract}

\maketitle

\section{Introduction}
Fundamental interactions of known elementary particles are well described by the the Standard Model~(SM) of particle physics.
Nevertheless, the SM cannot be the complete description of Nature since it does not account for neutrino oscillations, does not provide a viable dark matter candidate and cannot explain the baryon asymmetry of the Universe. Moreover, the SM suffers from several hierarchy problems, such as the stability of the Higgs mass to quantum corrections and the strong~CP problem. In addition, it contains intriguing puzzles related to the observed large hierarchies in the charged fermion masses and quark mixing angles. However, non of these pending issues which call for new physics~(NP) beyond the SM 
indicate a specific energy scale at which the associated NP phenomena will manifest themselves. Hence, a broad experimental program must be pursued.

While accelator-based experiments search directly for new particles over many orders of the mass and couplings, precision low-energy measurements may also reveal indirectly in a complementary approach the existence of new phenomena. Precise atomic physics table-top experiments are promising in this regard.
For example, observables which violate discrete symmetries of QED, as in parity-violating transitions~\cite{Porsev:2009pr,Tsigutkin:2009zz,PhysRevA.81.032114,Leefer:2014tga,Versolato:2011zza,Dzuba:2012kx,Roberts:2014bka,Jungmann:2014kia}, are a powerful tool to probe NP~\cite{Agashe:2014kda}.

Parity-conserving transitions are also interesting probes of NP which, however, require higher theoretical control.
Frequency measurements of narrow atomic transitions in heavy elements are possible within a few$\times10^{-16}$ relative accuracy~\cite{PhysRevLett.113.210801,PhysRevLett.108.090801} thanks to the optical frequency comb technique~\cite{PhysRevLett.84.3232,REICHERT199959}. Moreover, the experimental uncertainty in some systems now reaches the $10^{-18}$ level~\cite{PhysRevLett.116.063001,Bloom:2013uoa}, thus indicating the possibility of significant improvement in the near future. A limitation in translating the experimental precision to a bound on NP is the theory uncertainty, which is by far larger than the experimental precision. However, the high accuracy achieved in frequency measurements of narrow atomic transitions in heavy elements can be exploited to probe NP provided new observables largely insensitive to theory uncertainties are identified. For example, Refs.~\cite{Delaunay:2016brc,Frugiuele:2016rii,Delaunay:2016zmu,Berengut:2017zuo} proposed to bound new interactions between neutrons and electrons by testing linearity of King plots~\cite{King:63} of isotope shift~(IS) measurements. 

The situation is different for atoms and ions with few electrons and a small number of nucleons. There, QED calculations are carried out to high accuracy. 
For instance, low excited states in helium can be calculated up to $\mathcal{O}(\alpha^6)$ corrections, which yields theoretical predictions often more accurate than the measurements, see \textit{e.g.} Refs.~\cite{Pachucki:2017xcg,Karshenboim:2000kv}. In addition, nuclear finite size effects are also well described thanks to the extraction of the nuclear radius  from electron scattering experiments and muonic atom spectroscopy.
This allows to directly compare  theory and experiment in order to probe NP interactions~\cite{Karshenboim:2010ck,Karshenboim:2010cg,Jaeckel:2010xx,Ficek:2016qwp}. 

In this work we derive new limits on spin-independent NP interactions between the proton/neutron and the electron from optical frequency measurements, as well as from frequency shifts between different isotopes,  in hydrogen and helium atoms, helium-like ions such as lithium and nitrogen. In addition, we constrain new electron-electron interactions via precision spectroscopy of helium and positronium atoms.
We compare the resulting bounds on the product of the electron and neutron couplings and on the electron coupling separately, with existing indirect constraints and future prospects from IS measurement in heavier systems with many electrons. Furthermore, we perform global fits using all available transitions to constrain simultaneously the electron, neutron and proton couplings in a model-independent way. Finally, as a phenomenological application, we interpret the bounds within the dark photon model.

\section{Constraining new spin-independent interactions in atoms} 
\label{sec:NPatoms}

Consider a new force mediated by a boson $\phi$ of mass $m_\phi$ and spin $s=0,1,2$ with spin-independent couplings to the electron, proton and neutron $y_e$, $y_p$ and $y_n$, respectively.
At atomic energies, the exchange of $\phi$ is described by an effective potential between a nucleus $N$ of charge $Z$ and mass number $A$ and its bound electrons as
\begin{align}
	\label{eq:Ven}
	V_{N}(r) =  \frac{(-1)^{s+1}}{4\pi}y_e y_N \frac{e^{-m_\phi r}}{r} \, ,
\end{align}
where $y_N \equiv  y_pZ + (A-Z)y_n$ and $r$ is the electron-nucleus distance. 

For atoms with more than one electron, like helium, an effective potential between electron pairs is also induced  
\begin{align}
	\label{eq:Ve}
	V_{e}(r_{12}) = (-1)^{s+1} \frac{y_e^2}{4\pi} \frac{e^{-m_\phi r_{12}}}{r_{12}} \, ,
\end{align}
where $r_{12}\equiv |\vec{r}_1-\vec{r}_2|$ is the distance between electron~$1$ and electron~$2$, with $\vec{r_1}$, $\vec{r_2}$ describing their positions relative to the nucleus. 

The total frequency shift induced by the above potentials for a transition $i$ between atomic states $a$ and $b$ (with $E_b>E_a$) is described by first-order perturbation theory as
\begin{align}
\delta_{\rm NP}\nu^A_i 
	=(-1)^{s+1}\left(y_ey_N X_i+y_e^2 Y_i\right)\,,\label{eq:NPcont}
\end{align}
where $X_i$ and $Y_i$ are overlap integrals with the electronic wavefunctions depending only on the $\phi$ boson mass. In this paper, we focus on electronic transitions in helium-like and hydrogen atoms, for which the $X_i$ and $Y_i$ functions are calculated within first-order perturbation theory using non-relativistic wavefunctions as detailed in Appendices~\ref{app:HeWF} and~\ref{app:XYHe}.
While QED calculations rely on much more sophisticated wavefunctions, the use of the non-relativistic ones is a sufficient approximation to the dominant NP effects. Helium wavefunctions are very well approximated by antisymmetrized combinations of hydrogenic wavefunctions with effective nuclear charges accounting for the electronic screening~\cite{Eckart}, except for the $2S$ spin-singlet state where an accurate description of the inter-electron repulsion requires the use of Hylleraas functions~\cite{Hylleraas1928,Hylleraas1929}.

Taking into account the NP contribution in Eq.~\eqref{eq:NPcont}, the theory prediction for the frequency of an electronic transition $i$ in an isotope $A$ is given by
\begin{align}
	\label{eq:nuAi}
	\nu^A_{i} = 
	&\nu^{A}_{i,0} + F_i \langle r^2\rangle_A + \delta_{\rm NP}\nu^A_i \, ,  
\end{align}
where $\nu^A_{i,0}$ is the dominant contribution calculated in the point-like nucleus limit (including spin effects and nuclear polarizability), whereas the second term describes the leading finite nuclear size effects, $\langle r^2\rangle_A$ being the nuclear charge radius squared and $F_i$ is the field-shift~(FS) constant.
%
The IS between two isotopes $A$ and $A'$ for this transition is then described as 
\begin{align}
	\label{eq:nuAA}
	\nu^{A,A'}_i 
=	\nu^{A,A'}_{i,0} + F_i \delr_{A,A'} + y_ey_n X_i (A-A') \,,
\end{align}
where $\nu^{A,A'}_i\equiv \nu^A_i-\nu_i^{A'}$ and $\delr_{A,A'}\equiv\langle r^2\rangle_A-\langle r^2\rangle_{A'}$.
Higher-order effects of nuclear charge radius and finite magnetic radius 
are not resolvable by the current experimental accuracy~\cite{Pachucki:2017xcg,2015JPCRD..44c1206P}
and are therefore omitted in Eqs.~(\ref{eq:nuAi}) and (\ref{eq:nuAA}).

Absolute frequencies and IS in hydrogen and helium are calculated to high accuracy in the limit of point-like nuclei. 
However, the full theoretical prediction is often limited by the uncertainty related to the finite nuclear size effects. 
For a comparison 
between the experimental value and the QED prediction in the point-nucleus limit, we define 
\begin{align}
	\label{eq:Deltadef}
	\Delta^A_i \equiv \nu_{i,{\rm exp}}^A - \nu_{i,{0}}^A \, .
\end{align}

The NP contribution generically depends on the three coupling constants, $y_e$, $y_p$, and $y_n$, and the mediator mass, $m_\phi$. At fixed $m_\phi$ the product $y_e y_n$ can be probed independently of $y_p$ from a single IS measurement  (for any transition $i$) using Eq.~\eqref{eq:nuAA} 
\begin{align}
	\label{eq:IS1transition}
	y_e y_n = \frac{\Delta^{A,A'}_i - F_i \delr_{A,A'}}{X_i(A-A')} \, ,
\end{align}
where $\Delta_i^{A,A'}\equiv \Delta_i^A - \Delta_i^{A'}$ using Eq.~(\ref{eq:Deltadef}).
Hence the NP bound depends on the change in the mean-square nuclear charge radius, $\delr_{A,A'}$, which is measured either in electron scattering or in muonic atom spectroscopy experiments. 
Whenever applicable, the latter typically yields much more precise values of the charge radii.
In principle, the charge radius determination via electron scattering  may be affected by NP. However, we find that NP is only noticeable there for large coupling values that are already excluded by more sensitive probes. Hence the charge radius extraction from electron scattering cannot  be contaminated by NP.
Muonic atom spectroscopy measurements are more sensitive to NP contributions, especially in the keV--MeV mass range, and existing constraints on the $ y_{\mu} y_n $ coupling product do not rule out the possibility of NP contaminations in this region~\cite{Liu:2016qwd,TuckerSmith:2010ra,Beltrami:1985dc}. Therefore, for simplicity we will henceforth assume $ y_{\mu}=0$.

Alternatively, the charge radius dependence can be eliminated using an IS measurement in a second transition, yielding 
\begin{align}
	\label{eq:IS2transition}
	y_e y_n 
=	\frac{F_2 \Delta_1^{A,A'}-F_1\Delta_2^{A,A'} }{ (F_2 X_1 - F_1 X_2) (A-A') } \, ,
\end{align}
which, besides $X_{1,2}$, depends only on quantities known theoretically with high accuracy, namely $F_2/F_1$ and $\nu_{i,0}^{A,A'}$. The main drawback of Eq.~\eqref{eq:IS2transition} relative to Eq.~\eqref{eq:IS1transition} is the possible loss of sensitivity when $(F_2X_1-F_1X_2)\to 0$~\cite{Delaunay:2016brc,Berengut:2017zuo}. 
The latter is rather severe {\it for all} $m_\phi$ when close-by transitions with $F_1\approx F_2$ and $X_1\approx X_2$ are used, for example in transitions involving different states of the same fine-structure multiplet.
Another disadvantage of Eq.~\eqref{eq:IS2transition} is represented by how rapidly the sensitivity to $y_ey_n$ weakens at large mass. While $X_i\propto m_\phi^{-2}$ above $m_\phi\sim\mathcal{O}(10\,$keV), this leading term cancels out in the difference $(F_2X_1-F_1X_2)\propto m_\phi^{-3}$~\cite{Berengut:2017zuo}. Equation~\eqref{eq:IS2transition} bears resemblance to the method proposed in Ref.~\cite{Berengut:2017zuo} for heavy elements. However, there, the less accurate theory calculations for $\nu_{i,0}^{A,A'}$ are traded for IS measurements between two additional isotope pairs.

NP contributions to the electron-electron and electron-proton interactions cancel out  to first approximation in the IS. Thus, they can be probed more efficiently in absolute frequency measurements, despite the lower absolute accuracy. Helium transitions are sensitive to both kinds of interaction. In fact a combination of two frequency measurements can resolve the $y_ey_N$ and $y_e^2$ coupling products, thanks to transition-dependent $X_i,Y_i$ constants in Eq.~\eqref{eq:NPcont}. 
Hydrogen frequencies are sensitive to electron-proton interactions and have been used previously to probe $y_e y_p$ \cite{Karshenboim:2010ck,Karshenboim:2010cg,Jaeckel:2010xx}. However, presently unresolved issues related to the well-known proton radius puzzle \cite{protonsize,protonsize2} limits the application of hydrogen spectroscopy for constraining new atomic forces. 
Therefore, we will not use absolute spectroscopy measurements of hydrogen or deuterium as a probe of NP. 
Instead, we will consider hydrogen-deuterium IS spectroscopy since in this case there is no tension between $\delr$ values extracted from electronic and muonic measurements~\cite{Pohl1:2016xoo}.

\section{Bounds from Isotope shift measurements }
\label{sec:ISbound}
\begin{figure}[!t]
\begin{center}
\includegraphics[width=\columnwidth]{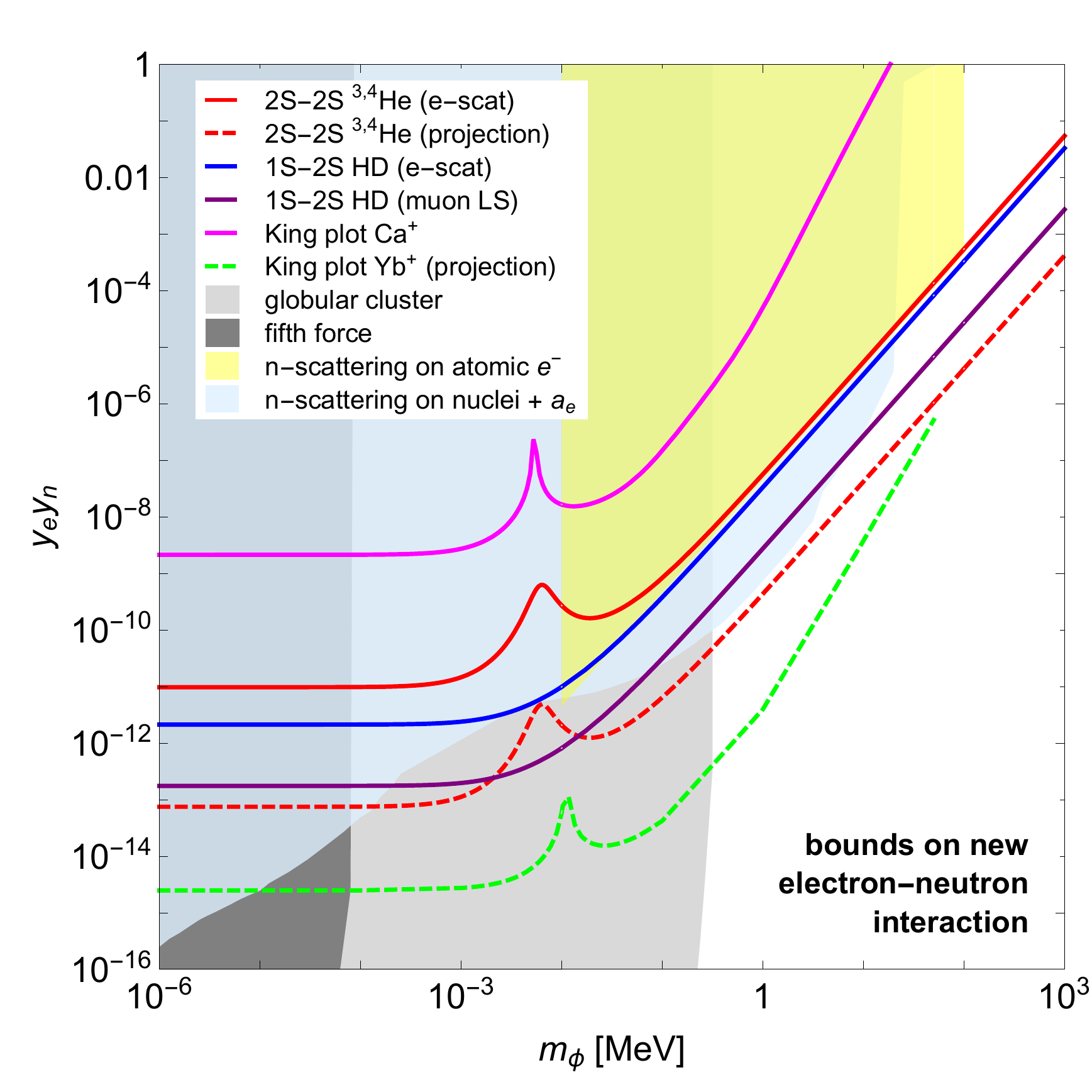}
\caption{Summary of the indirect constraints on a new electron-neutron interaction: isotope shift spectroscopy of helium and hydrogen/deuterium (this work) with the charge radius determined from electron scattering (``e-scat") or Lamb shift in muonic atoms (``muon LS"); comparison with King plot analyses of IS in heavy atoms (Ca$^+$, Yb$^+$)~\cite{PhysRevLett.115.053003,Berengut:2017zuo}, fifth force experiments~\cite{Bordag:2001qi,bordag2009advances}, electron-neutron scattering~\cite{Adler:1974ge}, neutron-nucleus scattering~\cite{Barbieri:1975xy,Leeb:1992qf,Nesvizhevsky:2007by,Pokotilovski:2006up} combined~\cite{Berengut:2017zuo} with $a_e$, and globular cluster~\cite{Hardy:2016kme}. The existing bounds are in solid lines, while the projections are in dashed lines. The He projection assumes a combined theory and experimental uncertainty of 100\,Hz; the Yb$^+$ projection assumes King linearity at 1\,Hz.  
}
\label{fig:yeynmphiall}
\end{center}
\end{figure}

Let us first discuss probes of new electron-neutron interactions. We focus here on spectroscopic probes based on IS measurements in helium, helium-like and hydrogen/deuterium atoms. The comparison of theory to experiment directly probes $y_e y_n$  independently of the presence of a NP coupling to protons. As shown in Table~\ref{tab:He34NPmax} (a full list of our input values is given in Appendix~\ref{app:data}), the theory uncertainty (in the point-like nucleus limit) is currently smaller than the experimental error for transitions involving low excited states, and the sensitivity to NP is limited by the experimental determination of charge radius differences. Our results are summarized in Fig.~\ref{fig:yeynmphiall}  which show the best constraints on $y_e y_n $, as a function of the mediator mass, arising from transitions in hydrogen and helium(-like) atoms. 

For comparison, we also show constraints derived from King linearity in heavy atoms~\cite{Berengut:2017zuo} and from other (non-atomic) observables.
Those include laboratory constraints resulting from the anomalous magnetic moment of the electron $a_e\equiv(g-2)_e/2$~\cite{Olive:2016xmw,Hanneke:2010au}, neutron scattering on atomic electrons~\cite{Adler:1974ge} or nuclei~\cite{Barbieri:1975xy,Leeb:1992qf,Nesvizhevsky:2007by,Pokotilovski:2006up} and fifth force experiments~\cite{Bordag:2001qi,bordag2009advances}, as well as astrophysical constraints from SN\,1987a~\cite{Raffelt:2012sp} and globular clusters~~\cite{Yao:2006px, Grifols:1986fc, Grifols:1988fv, Hardy:2016kme, Redondo:2013lna}. Note that also spectroscopy in molecular ions and antiprotonic helium constrains spin-independent interactions of nucleons~\cite{Salumbides:2013dua,Ubachs:2015fuf,Biesheuvel2016,UBACHS20161}, but it results in weaker bounds than from neutron scattering and is therefore not shown in Fig.~\ref{fig:yeynmphiall}. 
Some of these constraints can be evaded in specific models~\cite{Feldman:2006wg,Nelson:2008tn,Burrage:2007ew,Burrage:2016bwy,Brax:2010gp}.
We notice that bounds from few-electron atoms provide the strongest (indirect) constraints in the region above 300\,keV where astrophysical bounds lose sensitivity. Note, however, that direct constraints (not shown here) also exist, which are particularly sensitive for $m_\phi>1\,$MeV. Yet, they strongly depend on the assumed branching ratios for relevant $\phi$ decay modes, see {\it e.g.}  Ref.~\cite{Alexander:2016aln} for a review.
In the near future a higher sensitivity to NP is still expected in the King linearity test of Yb$^+$~\cite{Berengut:2017zuo} compared to few-electron atom spectroscopy, despite the projected improvements in helium transitions. 
In the following subsections, we discuss in detail the bounds obtained from IS measurements in helium, helium-like ions and hydrogen/deuterium atoms. 

\subsection{Helium and helium-like isotope shifts}
\label{sec:ISHelike}

\begin{figure}[!t]
\begin{center}
\includegraphics[width=\columnwidth]{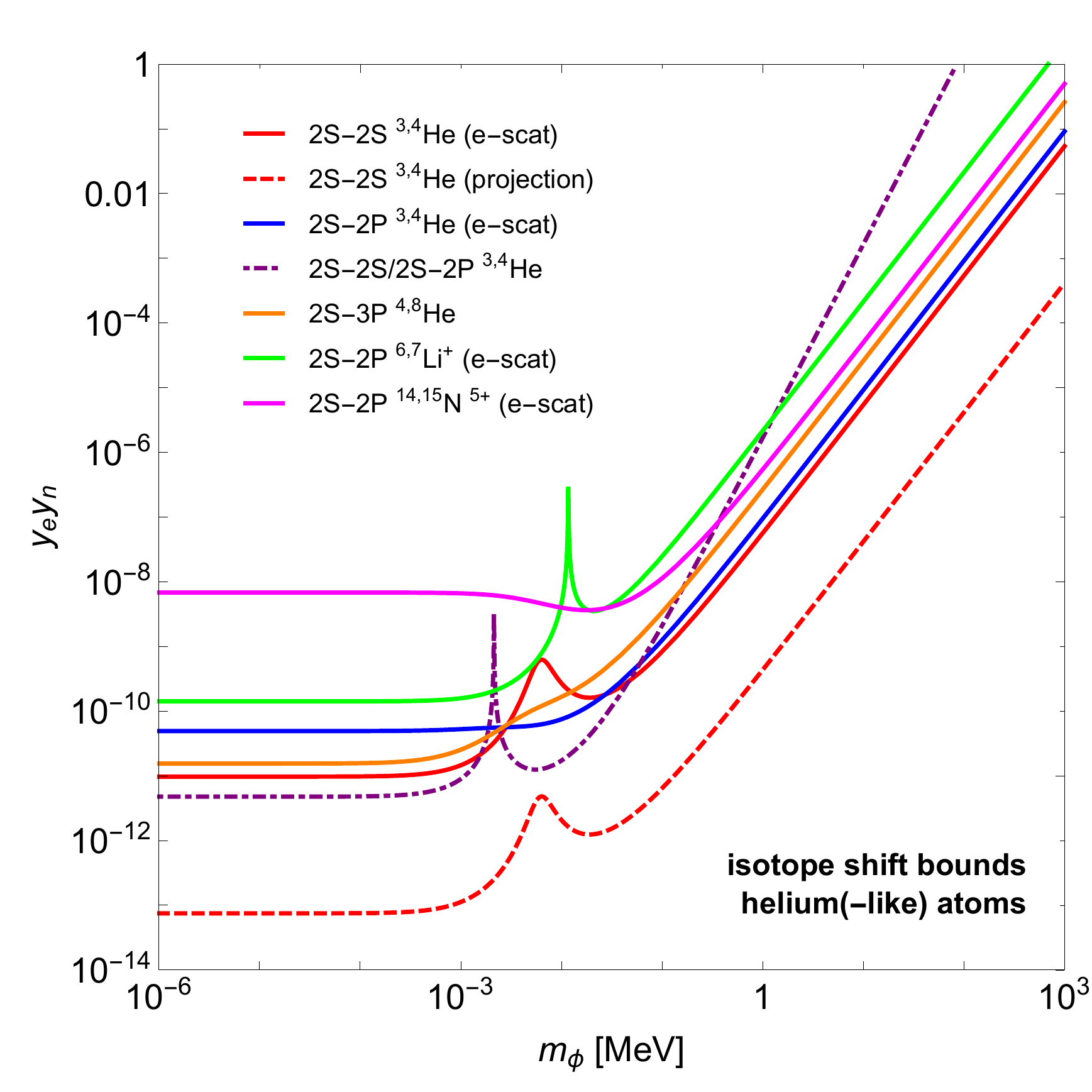}
\caption{Isotope shift bounds on new electron-neutron interaction from helium spectroscopy. The solid red and blue lines are the limits from the 1557$\,$nm and $1083\,$nm transitions, respectively, using the charge radii from e-He scattering. The dotted-dashed purple line is an illustration of the potential limit obtained by combining the two transitions (due to a $4 \sigma$ tension, see text for details).
The solid orange line is the upper bound derived from IS in unstable isotopes. 
The dashed red line represents the projected sensitivity with the 1557$\,$nm transition using charge radii from muonic helium and assuming a combined experimental and theory uncertainty of 100$\,$Hz.}
\label{fig:yeynmphiHe34}
\end{center}
\end{figure}

We derive here IS bounds using precision spectroscopy in two-electron atoms. This includes constraints from measurements in helium and helium-like lithium and nitrogen ions, all of which are presented in  Fig.~\ref{fig:yeynmphiHe34}, while the strongest one is also reported in Fig.~\ref{fig:yeynmphiall} for comparison with constraints from other atoms and different sources.

The most accurate IS in helium are measured within a few kHz uncertainty between $A=3,4$ for the $2^1S-2^3S$~\cite{Rooij} and $2^3P-2^3S$~\cite{Cancio1,Cancio2} transitions around $1557\,$nm and $1083\,$nm, respectively. While QED calculations in the point-nucleus limit reached sub-kHz accuracy, the theory prediction for the IS is limited by the charge radius difference $\delr_{3,4}$~\cite{Pachucki:2017xcg}. The latter can be extracted within a few percent from e-He scattering data~\cite{Sick:2015spa},
\begin{align}
	\label{eq:delrHe34}
	\delr_{3,4}^{\rm e-scat} = ( 1.067 \pm 0.065 ) \, {\rm fm}^2 \,.
\end{align}
Using Eq.~\eqref{eq:delrHe34} as an input for the theory predictions of He IS yields a good agreement between theory and experiment for both transitions, thus allowing to constrain NP electron-neutron interactions.

A higher sensitivity could be reached by combining the two transitions in order to eliminate $\delr_{3,4}$. In that case, Eq.~\eqref{eq:IS2transition} results in
\begin{align}
	\label{eq:He34twotrans}
	y_ey_n 
\approx 	 \frac{(-51\pm 14)\, {\rm kHz}}{ ( 5.7X_{1557}-X_{1083})  }\,,
\end{align}
which is $\sim 4 \sigma\,$ away from zero. 
Thus, it is not justified, given such a disagreement, to use the above to set limits on NP. 
Note, however, that this large deviation is the mere consequence of a known tension between the two transitions which 
may originate from underestimated uncertainties \cite{Pachucki:2017xcg}. 
Despite this circumstance, it remains interesting to observe that in the case that the tension will be resolved by refined QED calculations and/or measurements, the expected sensitivity to $y_e y_n$ is stronger by a factor $\sim6$ relative to the use of $\delr_{3,4}^{\rm e-scat}$. In the (yet implausible) event that the above deviation is an evidence for a new electron-neutron interaction, the latter should be visible in other atomic systems. For instance, Eq.~\eqref{eq:He34twotrans} would imply a violation of King linearity in ytterbium ion clock transitions at the $\mathcal{O}(100\,$Hz) level~\cite{Berengut:2017zuo}. 

Alternatively, $\delr_{3,4}$ can be extracted with high accuracy from muonic helium spectroscopy. The CREMA collaboration is currently conducting Lamb shift measurements in muonic He$^+$ aiming at a determination of $^{3,4}$He charge radii with a relative uncertainty of $3\times 10^{-4}$~\cite{Antognini:2011zz}. Assuming this will result in a $\delr_{3,4}$ value consistent with e-He scattering and (electronic) helium spectroscopic data, the sensitivity to NP will hence be limited by the experimental accuracy in helium IS measurements. 
Moreover, future IS measurements in the $2^1S-2^3S$ transition down to $\mathcal{O}(100\,$Hz) precision are expected~\cite{VassenTalk:2017}, with a comparable theory improvement. Hence, this would potentially improve sensitivity to NP effects for that transition by two orders of magnitude. As shown in Fig.~\ref{fig:yeynmphiall}, this is still weaker than the sensitivity expected from King linearity violation in ytterbium ions, except for $m_\phi \gtrsim 10\,$MeV due to the different scaling of the bound with the mediator mass ($m_\phi^{2}$ versus $ m_{\phi}^{3}$).\\

Precision measurements are also achievable in heavier (unstable) helium isotopes. For instance, IS between $A=4$ and $A=6,8$ isotopes for the $2^3S-3^3P$ transition ($389\,$nm) are measured with $\sim\!100\,$kHz accuracy~\cite{Mueller:2008bj}. However, the situation is different here since there is no independent measurement of the $^{6,8}$He charge radii and the FS cannot be reliably predicted for the $389\,$nm transition. Nevertheless one can still derive an upper bound on NP by saturating the difference between theory (assuming a point-like nucleus) and experiment, which corresponds to setting $\delr_{AA'}=0$ in Eq.~\eqref{eq:IS1transition}. Since $\Delta_{389}^{8,4}=-0.918\,$MHz~\cite{Mueller:2008bj}, the NP contribution is not strongly constrained. Yet, the resulting bound on $y_ey_n$ is strengthened by a factor of $A-A'=4$ which makes it comparable to the IS bound from the $1083\,$nm transition. An order of magnitude improvement could be obtained with an independent determination of the charge radii of $A=6,8$ isotope of helium.\\

Finally, IS in helium-like ions are also well measured. The highest accuracy is obtained in singly-ionized lithium~\cite{PhysRevA.49.207} and five-times ionized nitrogen~\cite{PhysRevA.57.180}. The measured frequency shifts are between $A=6,7$ in the $2^3S-2^3P$ transition for Li$^+$, and between $A=14,15$ in the $2^1S-2^3P$ transition for N$^{5+}$. We rely on the theory predictions used in the quoted references. This assumes nuclear charge radii determined from electron-scattering data~\cite{JagerVries} with a relative accuracy of $\sim2\,\%$  for lithium and from electron-scattering data~\cite{DEVRIES198822} and muonic X-ray line measurements~\cite{SCHALLER1980333} with $\sim0.5\,\%$ for nitrogen. The resulting bounds are weaker than the ones from helium. 
Further precision measurements with helium-like boron and carbon ions are also underway~\cite{Pachucki:2017xcg}. 

\begin{table}[!t]
\begin{center}
\begin{tabular}{cccccc}
\hline\hline\noalign{\smallskip}
isotopes & transition & $\delta_{\rm NP}\nu$ & $\sigma_{\nu_{\rm exp}}$ & $\sigma_{\nu_{0}}$ & $\sigma_{\delr}$ \\
\noalign{\smallskip}\hline\noalign{\smallskip}
$^{3}$He/$^{4}$He & \begin{tabular}{c} $2^1S-2^3S$ \\ $2^3P-2S$\end{tabular} & \begin{tabular}{c} $+9\pm14$ \\ $-2\pm78$ \end{tabular} & \begin{tabular}{c} $2.4$ \\ $3.3 $ \end{tabular} & \begin{tabular}{c} $0.19$ \\  $0.9$ \end{tabular} 
& \begin{tabular}{c} $14$ \\ $78$\end{tabular}  \\
\noalign{\smallskip}\hline\noalign{\smallskip}
H/D& \begin{tabular}{c} $1S-2S$ \\ $2S-12D$\end{tabular}& \begin{tabular}{c} $+76\pm61$ \\ $+1.2\pm 10$ \end{tabular} & \begin{tabular}{c} 0.02 \\ 
9.3 \end{tabular} & \begin{tabular}{c} 0.9 \\ 4.2\end{tabular} 
 &\begin{tabular}{c} 61 \\ $ $ \end{tabular} \\
\noalign{\smallskip}\hline\hline
\end{tabular}
\end{center}
\caption{Allowed NP contributions $\delta_{\rm NP}\nu$ for the most accurate IS measurements in helium and hydrogen isotopes, along with the standard uncertainties from experiment, $\sigma_{\nu_{\rm exp} }$, QED calculation (point-nucleus limit, $\sigma_{\nu_{0}}$) and charge radius difference extracted from electron-scattering data, $\sigma_{\delr}$. 
Only for $2S-12D$ in H/D, $\sigma_{\nu_0}$ refers to the complete theory prediction including the FS. All numbers are in kHz. For references and input values, see Tables~\ref{tab:inputIS} and \ref{tab:absfreq}.}
\label{tab:He34NPmax}
\end{table}

\subsection{Hydrogen-deuterium shifts}
\label{sec:ISHD}
Hydrogen-deuterium shifts are complementary probes of new electron-neutron interactions.
The most accurate IS measurement is for the $1S-2S$ transition ($121.6\,$nm), with $\sim10^{-11}$ relative uncertainty~\cite{Parthey:2010aya,PhysRevA.83.042505}. The QED calculation is less precise by a factor of $\sim60$, being equally limited by the experimental value of the proton-to-electron and deuteron-to-electron mass ratios as well as higher-order corrections to the Lamb shift and nuclear polarizability~\cite{Parthey:2010aya}. Additional IS measurements exist with lower precision, including the $2S-nS/D$ transition series for $n=8,12$ states~\cite{deBeauvoir2000,PhysRevLett.78.440,PhysRevLett.82.4960}, and the frequency differences~\cite{weitz1995precision}
\begin{align}
	\label{eq:nuLS}
	\nu_{\rm LS}^L\equiv \nu_{2S-4L}-\frac{1}{4}\nu_{1S-2S}\,,
\end{align}
with $L=S,D$. The latter is constructed such that the leading contribution from Coulomb-like potentials cancels out, thus making it directly sensitive to Lamb shift~(LS) corrections. As a result, $\nu_{\rm LS}^L$ becomes less sensitive to NP with an interaction range longer than the atomic size $\sim a_0=(\alpha m_e)^{-1} \approx (4\,$keV$)^{-1}$. 
Since all transitions in the $2S-nS/D$ series have comparable sensitivity to NP,  we consider only the $2S-12D$ transition for illustration.
  
Here again, the FS contributions are least known theoretically as they are limited by the charge radius difference $\delr_{2,1}$ between the deuteron and the proton. The latter can be extracted either from electron scattering data\footnote{We use here the proton radius value extracted from the so-called Mainz data~\cite{PhysRevC.90.015206}.}, which yields~\cite{Mohr:2012tt}
\begin{align}
	\label{eq:delr2HDscatt}
	\delr^{\rm e-scat}_{2,1} 
\!=\!  \left( 3.764\pm0.045 \right)\, {\rm fm}^2\,,
\end{align}
or muonic hydrogen/deuterium spectroscopy~\cite{Pohl1:2016xoo} 
\begin{align}
	\label{eq:muonRDp}
	\delr_{2,1}^{\mu}=  \left( 3.8112\pm0.0034 \right)\, {\rm fm}^2 \, .
\end{align}
Note that the charge radius differences in Eqs.~\eqref{eq:delr2HDscatt} and~\eqref{eq:muonRDp} are consistent within uncertainties, despite the (still puzzling) significant discrepancies between muonic and electronic determinations of the proton~\cite{protonsize2,protonsize} and deuteron~\cite{Pohl1:2016xoo} radii.
Using $\delr_{2,1}^\mu$ to predict the FS contribution yields a sensitivity to NP larger by a factor $\sim13$ relative to $\delr_{2,1}^{\rm e-scat}$, assuming the radii extraction from muonic spectroscopy is not affected by a possible NP coupling to muons.

\begin{figure}[!t]
\begin{center}
\includegraphics[width=\columnwidth]{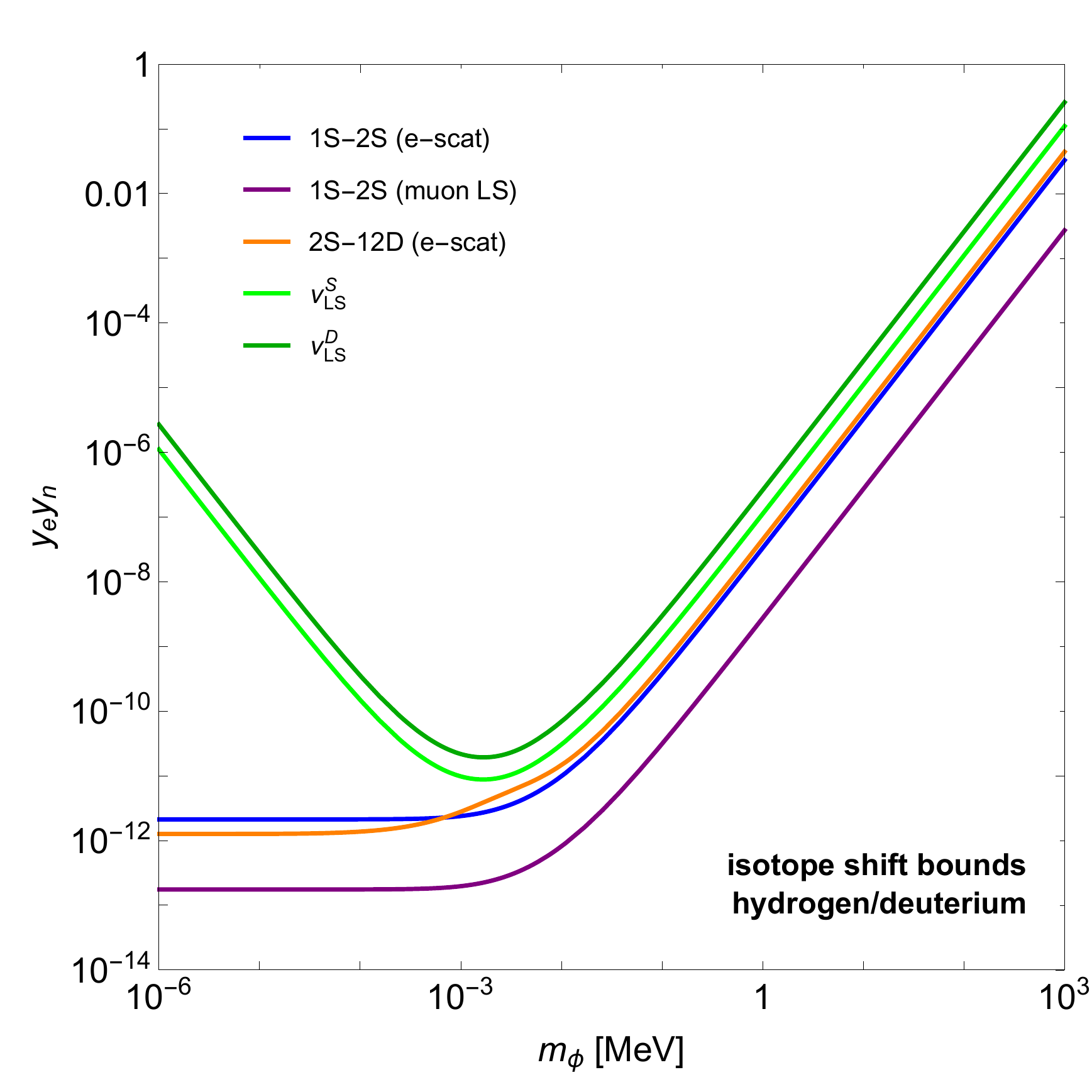}
\caption{Isotope shift bounds on new electron-neutron interaction from hydrogen/deuterium spectroscopy. The blue (purple) line is the limit from the $1S-2S$ transition using charge radii from electron scattering (muonic spectroscopy). The orange and green lines are limits derived from the $2S-12D$ transition and the $\nu_{\rm LS}^{S,D}$ observables, respectively, using electron-scattering data. See text for details.}
\label{fig:yeynmphiHD}
\end{center}
\end{figure}

The IS bounds on a new electron-neutron interaction from hydrogen/deuterium are summarized in Fig.~\ref{fig:yeynmphiHD}.

\section{Bounds from absolute frequency measurements}
\label{sec:absfreq}

While IS are only sensitive to electron-neutron interactions, absolute frequencies can also probe the electron-proton and, in atoms with more than one electron or in positronium, electron-electron interactions.
As we discussed above, by measuring two transitions one can extract $y_e$ and $y_N$ separately, and the combination with the IS data will also allow for a separation of $y_p$ from $y_n$.   

In case $\phi$ couples both to protons and neutrons with a similar strength, as in a Higgs portal or gauged $B-L$, the sensitivity to probe NP with IS is expected to be stronger than from the absolute frequency measurements. This can be understood as follows. In light atoms, the NP contributions to the IS and to the absolute frequency are of the same order. 
However, typically the absolute accuracy of IS data (theory and experiment) is better by at least an order of magnitude than the absolute frequency data, see Appendix~\ref{app:data}.  
Thus, for $y_n\approx y_p$ IS measurements are a more sensitive to NP than the absolute frequencies.  

It is important to distinguish between the case of a generic new force coupled to the electron and to the nucleus (including the proton) and the case of a dark photon (kinetic mixing) where the charges are proportional to the electric charges and a more careful treatment of the definition of the electromagnetic coupling, $\alpha$, is required, see Ref.~\cite{Jaeckel:2010xx}. 

\subsection{Bounds on $y_e$ from helium and positronium}
\label{sec:ye}

\begin{figure}[!t]
\begin{center}
\includegraphics[width=\columnwidth]{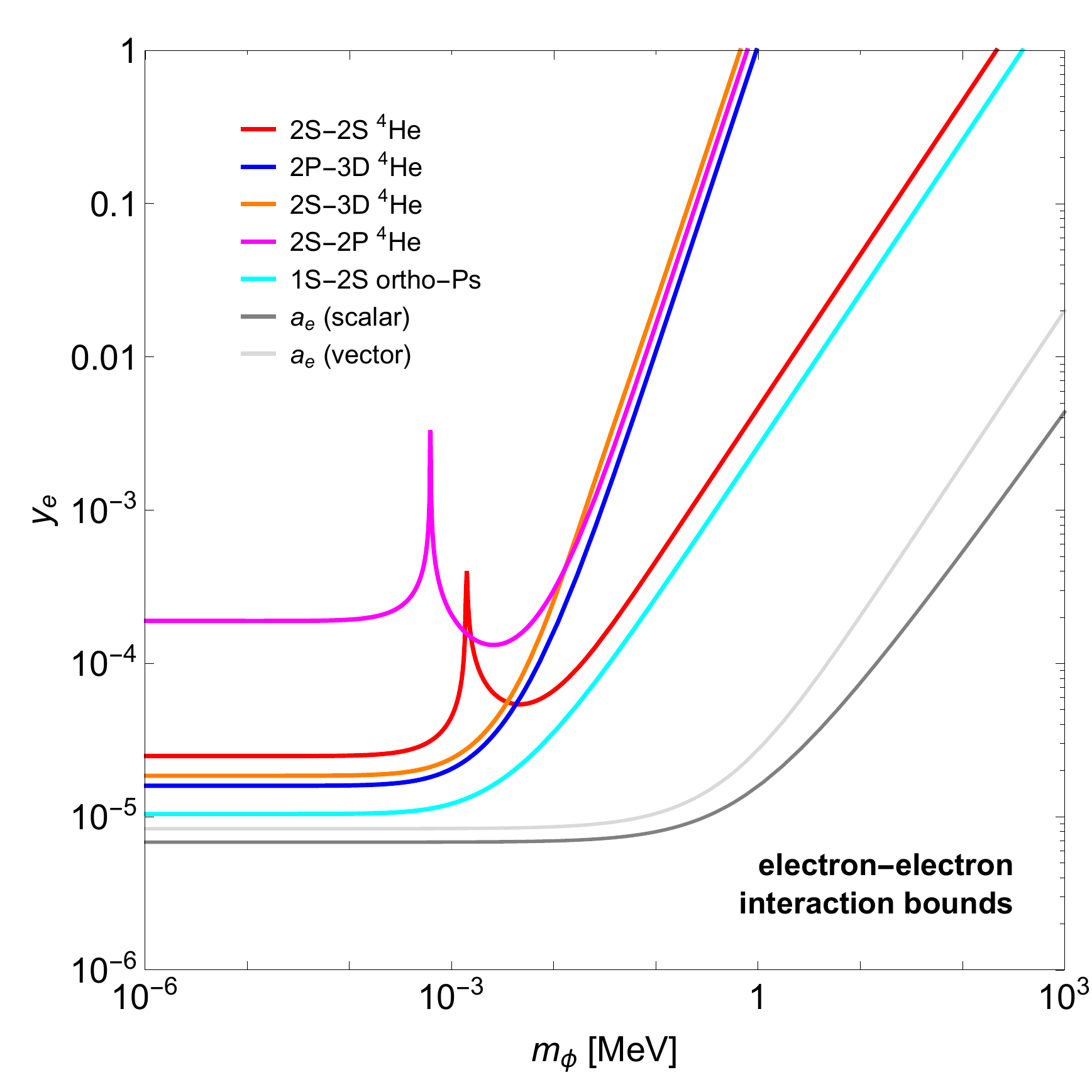}
\caption{The upper bound on $y_e$ as function of $m_\phi$ from $1^3S_1 - 2^3 S_1$ in positronium~\cite{Fee:1993zz,Czarnecki:1999mw}\,(cyan), helium data~\cite{Pachucki:2017xcg} and a comparison to electron magnetic moment, $a_e$, for a scalar (dark gray) and vector (light gray). }
\label{fig:ye}
\end{center}
\end{figure}

The electron-electron interaction can be probed in atoms with more than one electron, the simplest is helium, or in purely electronic systems such as positronium.
Starting with the positronium, the $1^3S_1 - 2^3 S_1$ interval is measured at the $10^{-9}$ level~\cite{Fee:1993zz} in a agreement with the theory prediction of Ref.~\cite{Czarnecki:1999mw}. 
For helium we combine all the transitions that are given in Table~II of Ref.~\cite{Pachucki:2017xcg}, where the agreement between theory and experiment is better than $2\sigma$; the full list is given in Appendix~\ref{app:data}. 
Thus, we use the above to put upper bounds on $y_e$ as function of the force-carrier mass $m_\phi\,$. The results are presented in Fig.~\ref{fig:ye}, where we also added the constraint from the electron magnetic moment, $a_e$, for comparison. This  shows that $a_e$ is still the strongest probe among the three. Yet the $\phi$ contribution to $a_e$ enters only at the loop level which makes it more prone to cancelation against additional contributions from other states present in a complete NP model. Note that the helium bounds in Fig.~\ref{fig:ye} are evaluated by assuming no electron-nucleus interactions. We have verified that marginalizing over the latter does not significantly change the bounds. 
The bounds from positronium and helium are comparable and below few keV are weaker than the bound from $a_e$ only a factor of few.

\begin{figure*}[!t]
\begin{center}
\includegraphics[width=8cm]{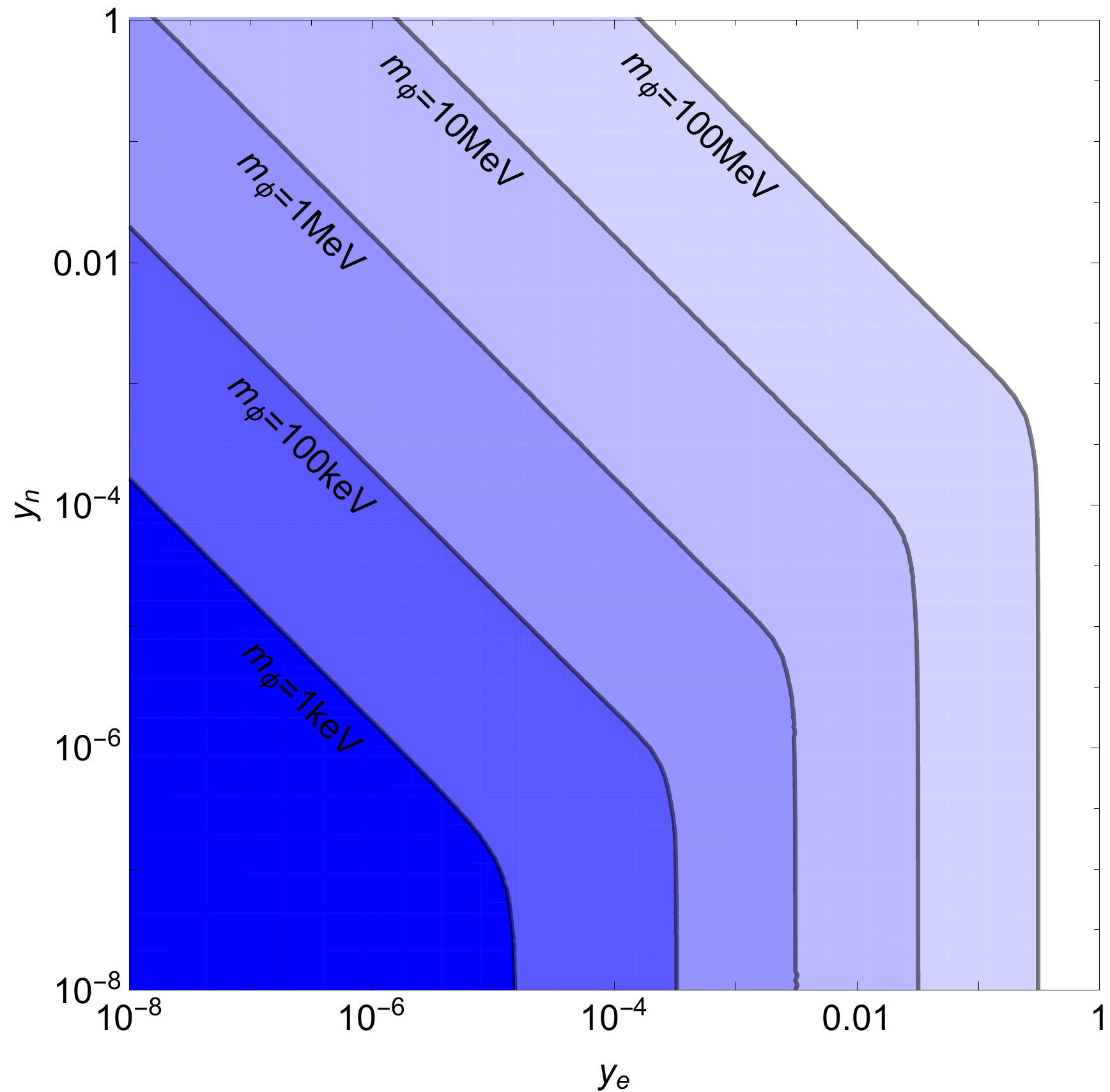}~~~~~~~~~~~~
\includegraphics[width=8cm]{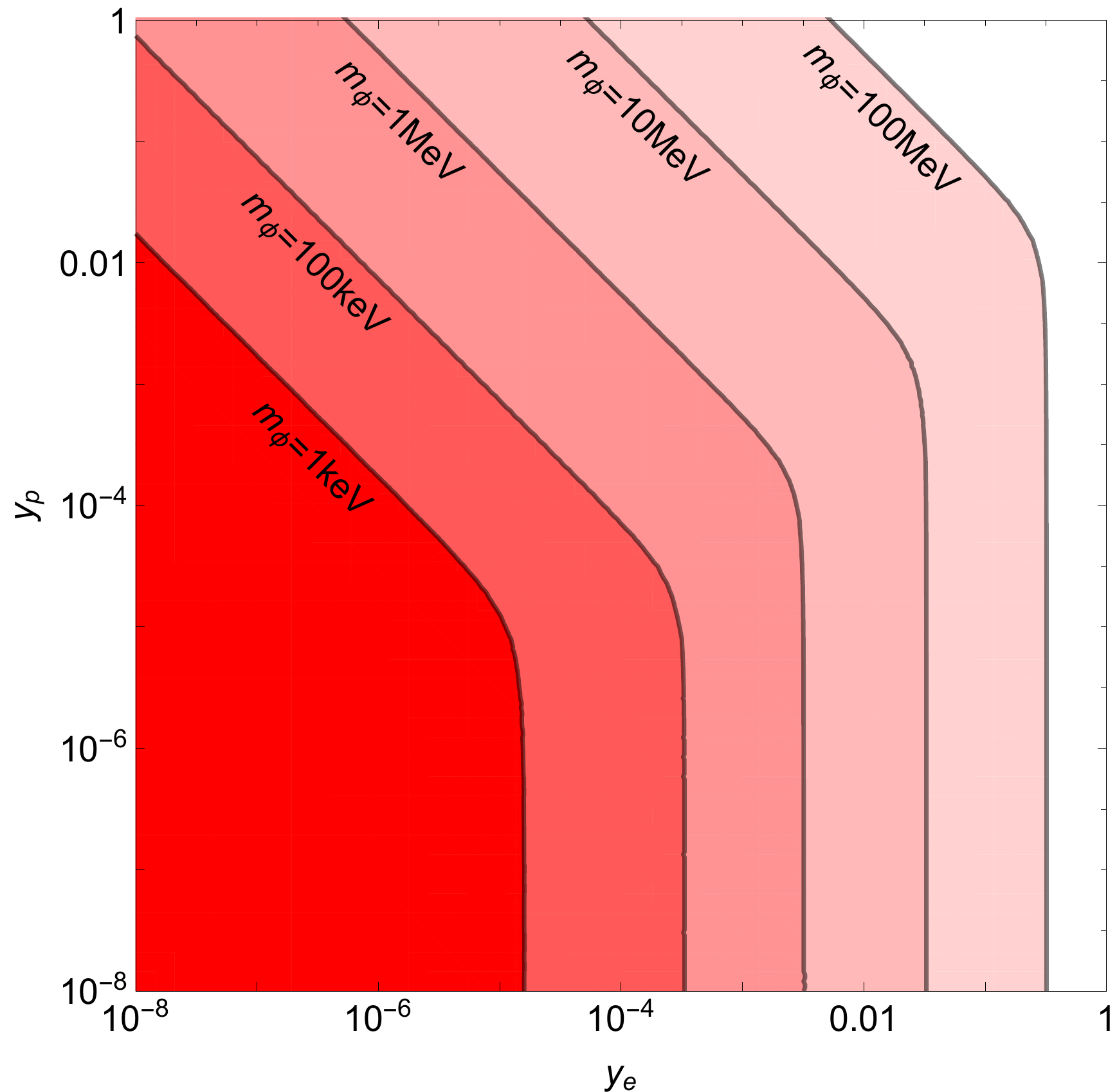}
\caption{The allowed region 95\,\%~CL contours for fixed mediator mass (as indicated in the plot) from the global $\chi^2$ analysis of helium and hydrogen absolute frequencies as well as their isotope shifts. The left (right) panel shows contours of $y_n$ ($y_p$) vs $y_e$, marginalizing over $y_p$ ($y_n$).}
\label{fig:chi2}
\end{center}
\end{figure*}

\subsection{Model-independent bounds on $y_e$, $y_p $ and $y_e $}
\label{sec:yeypHe}

Here we combine observables from different atoms to probe the NP couplings $y_p$, $y_n$ and $y_e$ independently. In order to do so we perform a global fit based on a $\chi^2$ function constructed from IS in hydrogen and helium as well as absolute transition frequencies in helium. 
Our $\chi^2$ is composed of the $2^1S-2^3S$ and $2^3P-2S$ IS between $^{3}$He and $^4$He, the helium/deuterium IS in the $1S-2S$ transition and the $\nu^{S}_{\rm LS}$ observable, and the absolute frequencies considered in Section~\ref{sec:ye}.
We present in Fig.~\ref{fig:chi2}, the 95\,\%~CL contours in the $y_e-y_p$ and $y_e-y_n$ planes for several values of $m_\phi$. For each pair of couplings, we marginalize over the third coupling $y_n$ and $y_p$, respectively.
The generic shape of the bounds is understood as follows. Since the overlap integrals $X_i,Y_i$ for electron-nucleus and electron-electron interactions are of comparable order when $y_{p,n}\gtrsim y_e$, absolute frequencies and IS constrain the products $y_e y_p$ and $y_e y_n$, respectively, leading to contours at 45 degrees. The latter are then truncated once $y_e$ reaches a large value (typically $y_e\gtrsim y_{p,n}$) so that the $y_e^2$ term dominates the NP contribution to absolute frequencies in helium and the bounds become independent of $y_p$ or $y_n$. Note that helium absolute frequencies are in principle sensitive to a possible relative sign between the nuclear and the electron couplings. We checked that either sign yields very similar bounds and thus, for simplicity, we  present global-fit results for positive couplings only.

\subsection{Atomic bounds on kinetic mixing}
\label{sec:DP}

\begin{figure}[!t]
\begin{center}
\includegraphics[width=\columnwidth]{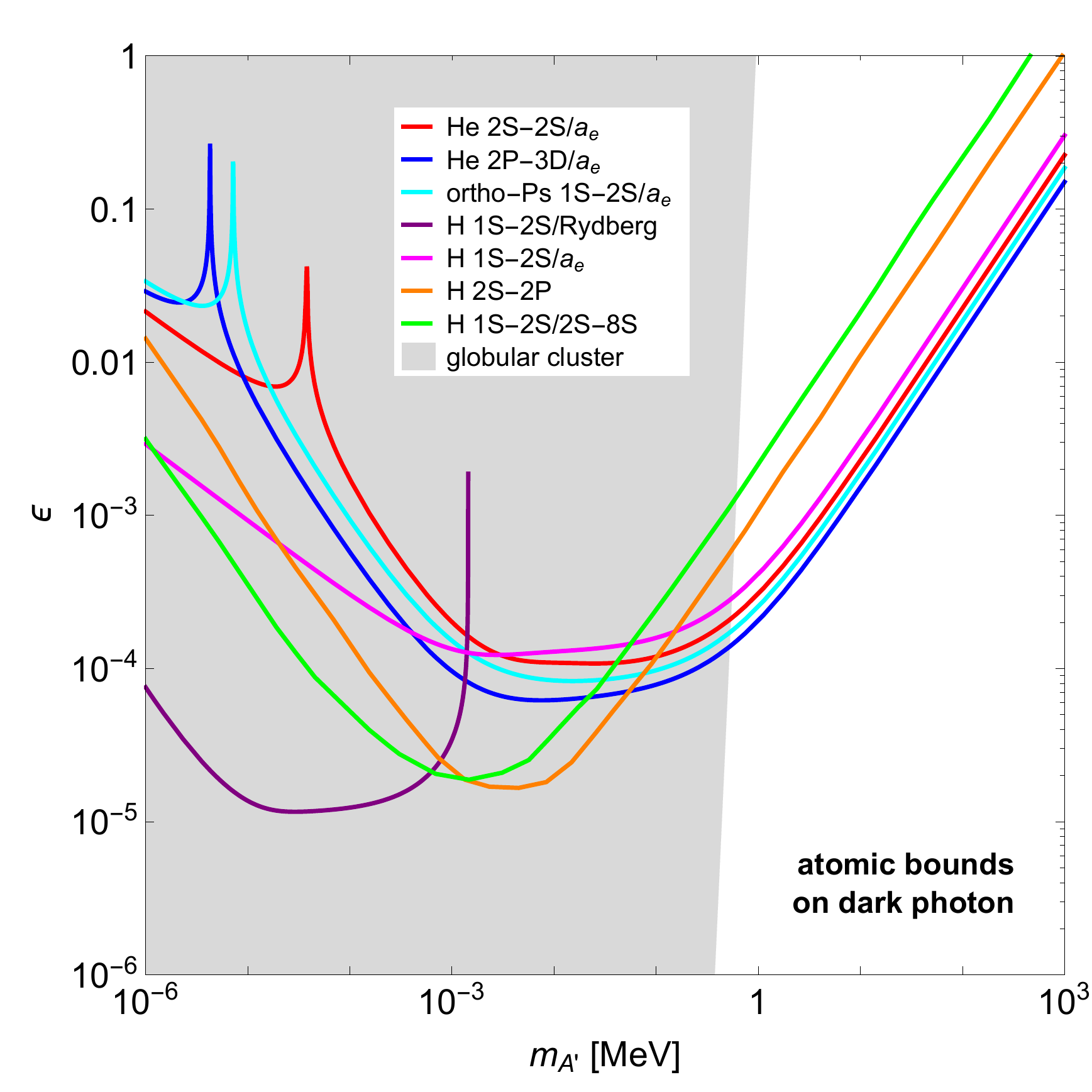}
\caption{Comparison of atomic (indirect) constraints on the mixing $\epsilon$ between the photon and the dark photon $A'$ depending on its mass $m_{A'}$. 
The red/blue (cyan) lines refer to He (positronium) transitions obtained in this work.
For comparison, previous bounds from H spectroscopy are shown for $1S-2S$ combined with Rydberg states (purple) or $a_e$ (pink)~\cite{Pospelov:2008zw,Karshenboim:2010ck,Karshenboim:2010cg} and from one or the combination of two H transitions (orange, green)~\cite{Jaeckel:2010xx}.
The range of $m_{A'} \leq 300$\,keV is excluded by globular clusters~\cite{Hardy:2016kme}; for $m_{A'}\gtrsim1$\,MeV several direct, decay-dependent bounds also apply.}
\label{fig:DP}
\end{center}
\end{figure}

For the sake of illustration, we apply now our result to a specific NP model, that of a kinetically mixed massive gauge boson, the dark photon, denoted as $A'$~\cite{Holdom:1985ag}. As a result of the mixing between the photon and the dark photon, $A'$ couples to the electromagnetic current, and its couplings to the protons, electrons and neutrons are $y_{p,n,e}=\epsilon e,0,-\epsilon e$, respectively, where $e$ is the QED gauge coupling constant and $\epsilon$ is a mixing parameter. 
Since all $A'$ couplings are determined by a single parameter, a single atomic transition would suffice to probe it. However, when $m_{A'}\lesssim 1/a_0$, the dark photon induces a $1/r$ atomic potential which is not distinguishable from the Coulomb one and the $A'$ effect is a mere redefinition of the fine-structure constant, $\alpha\to(1+\epsilon^2)\alpha$. Hence, in this regime, we need at least two observables to probe the dark photon, one of them being used to fix $\alpha$. 
We follow here the procedure of Ref.~\cite{Jaeckel:2010xx} and combine either two atomic transitions together or one transition with $a_e$, the anomalous magnetic moment of the electron. 
Figure~\ref{fig:DP} shows the $95\,\%$~CL bounds that we derived from helium and positronium, each combined with $a_e$, as well as 
existing bounds from hydrogen spectroscopy~\cite{Jaeckel:2010xx,Karshenboim:2010cg,Karshenboim:2010ck}. We find that helium and positronium bounds surpass the known hydrogen bounds above $\sim100\,$keV. We chose to present only indirect constraints from atomic spectroscopy on the kinetic-mixing parameter since those do not depend on the $A'$ decay mode.
In the sub-MeV region, these atomic probes are the most sensitive ones, after the LSND neutrino detector which directly searches for $A'$ in the 3 photons decay~\cite{Pospelov:2017kep}, and the study of star cooling in globular clusters which excludes, for $m_{A'}\lesssim 300$\,keV, mixing parameter values far below the displayed range of $\epsilon$ in Fig.~\ref{fig:DP}. 
For $m_{A'}\gtrsim1$\,MeV, the sensitivity of atomic spectroscopy is also much weaker compared to probes based on $A'$ decay (either visibly or invisibly) as in electron beam-dump experiments or colliders, like BaBar (see Ref.~\cite{Alexander:2016aln} for a review). In conclusion, for $ m_{A'}\gtrsim0.3$\,MeV, the most sensitive indirect probe of dark photon is from combining $a_e$ with atomic transitions in helium.

\section{Discussion}

In this work we study the sensitivity to new spin-independent forces of hydrogen and helium-like atoms considering both absolute frequency and isotope shift measurements.
We exploit the accuracy of both the measurements and the theoretical predictions achieved in these systems~\cite{Pachucki:2017xcg,Karshenboim:2000kv}.
We demonstrate for the first time the power of isotope shift measurements in few-electrons atoms to constrain models where the new degree of freedom, $\phi$, couples not only to the proton, but also to the neutron as, for instance, in the  $B-L$ and Higgs portal models.
The derived bounds represent, to date, the strongest laboratory bound on $y_e y_n$ for $ m_{\phi} \gtrsim 100$\,eV.
For masses heavier than 300\,keV, where astrophysical probes are ineffective, isotope shift spectroscopy in few-electron atoms constrains new regions of the parameter space in a  model-independent way. 
Previous works on spin-independent new interactions~\cite{Jaeckel:2010xx,Karshenboim:2010cg,Karshenboim:2010ck} focused on hydrogen which is only sensitive to new interactions between the electron and the proton. The highly precise spectroscopy of helium has the advantage to probe also the electron coupling alone, reaching  a sensitivity comparable to $a_e$ below few keV. (See Ref.~\cite{Ficek:2016qwp} for similar results regarding spin-dependent electron-electron interactions.)
Furthermore, we show that current  precision in positronium spectroscopy has comparable constraining power.

The present work emphasizes how the effort in improving the knowledge of the nuclear size has the indirect effect of improving the sensitivity to new spin-independent forces between the constituents of the atoms.

\begin{acknowledgments}
We thank  D.~ Budker, R.~Ozeri, K.~Pachucki and G.~Perez for useful discussions and careful reading of the manuscript.
We thank G.~B\'elanger, J.~Berengut, A.~Falkowski, O.~Hen, J.~Jaeckel and M.~Safranova for useful discussions and correspondence.
CD is supported by the program Initiative d'Excellence of Grenoble-Alpes University under grant Contract Number ANR-15-IDEX-02.
The work of YS is supported by the Office of High Energy Physics of U.S. Department of Energy~(DOE) under grant Contract Number  DE-SC0012567.
\end{acknowledgments}

\appendix

\section{Experimental Data and Theoretical Prediction} 
\label{app:data}

In this appendix we provide all experimental data and theoretical predictions that have been used in this paper.

We start with the nuclear charge radii in Tab.~\ref{tab:rNuc} used throughout the paper.

\renewcommand{\arraystretch}{1.2}
\begin{table}[h]
 \begin{center}
 \begin{tabular}{c c|c|c c}
\hline \hline
element	&$A$&$r_A$\,[fm]	&method	&ref.\\  \hline \hline
\multirow{4}{*}{H/D}
	&1	&0.8791 $\pm$ 0.0079	&\multirow{2}{*}{e-scat.}	&\cite{Mohr:2015ccw}
		\\
	&2	&2.130 $\pm$ 0.010	&	&\cite{Mohr:2015ccw}
		\\
	&1	&0.84087 $\pm$ 0.00039	&\multirow{2}{*}{$\mu$-spec.}	&\cite{protonsize,protonsize2}	\\
	&2	&2.12562 $\pm$ 0.00078	&	&\cite{Pohl1:2016xoo}	\\\hline
\multirow{2}{*}{He}
	&3	&1.973 $\pm$ 0.016	&\multirow{2}{*}{e-scat.}	&\cite{Sick:2015spa}
	\\
	&4	&1.681 $\pm$ 0.004	&	&\cite{Sick:2015spa}
	\\\hline
\multirow{2}{*}{Li}
&6	&2.589 $\pm$ 0.039		&\multirow{2}{*}{e-scat.}&\multirow{2}{*}{\cite{ANGELI201369} } \\
&7	&2.444 $\pm$ 0.042					&	& \\ \hline
\multirow{2}{*}{N}
&14	&2.560 $\pm$ 0.011	&\multirow{1}{*}{$\mu$-spec. } 	&{\cite{SCHALLER1980333}} 
\\
&15	&	2.612 $\pm$ 0.009	&	\multirow{1}{*}{e-scat. } 	&{\cite{DEVRIES198822}}			\\
\hline\hline
 \end{tabular}
 \caption{Charge radii $r_A$ used for the IS bounds for different elements with isotopes $A$, obtained via electron-scattering experiments (e-scat.) or spectroscopy in muonic atoms ($\mu$-spec.). 
}
 \label{tab:rNuc}
 \end{center}
\end{table}

In Tab.~\ref{tab:inputIS} we continue with the input values used for the IS bounds in Tab.~\ref{tab:He34NPmax} and Fig.~\ref{fig:yeynmphiHe34} of Sect.~\ref{sec:ISbound}, see also Tab.~5 of Ref.~\cite{Pachucki:2017xcg}.
Finally, we provide the values of measurements and calculations of absolute frequencies in Tab.~\ref{tab:absfreq}.

\renewcommand{\arraystretch}{1.5}
\begin{table*}[h!]
 \begin{center}
 \begin{tabular}{c|c||c c||c|c c}
 \hline\hline
 $A,A'$	&transition $i$	&$\nu_i^{AA',{\rm exp}}$\,[kHz]&ref. &$\nu_{i,0}^{AA',{\rm th}}$\,[kHz]&$F_i$\,[kHz/fm$^2$] &ref. \\ \hline\hline
 \multirow{2}{*}{$^{3,4}$He}
	&$2^1S^{1/2}-2^3S^{3/2}$&-8\,034\,286.259$\pm$2.4	&\multirow{2}{*}{\cite{drake2006atomic,PhysRev.187.5,PhysRevA.1.571}} 	&-8\,034\,065.91$\pm$0.19	&-214.66$\pm$0.02	&\multirow{2}{*}{\cite{PhysRevA.94.052508,Patkos:2017lnm,PhysRevA.95.012508,2015JPCRD..44c1206P} }
	\\
	&$2^3P-2S$		&-33\,668\,444.7$\pm$3.2	&	&-33\,667\,149.3$\pm$0.9	&-1\,212.2 $\pm$ 0.1	& \\ 
 $^{4,8}$He&	$2S-3P_2$		&64\,701\,466$\pm$52	&\cite{PhysRevLett.99.252501}
 	&64\,702\,409	&1\,008	&\cite{PhysRevLett.99.252501}
 	\\ \hline
 \multirow{1}{*}{H/D}
	&$1S-2S$		&670\,994\,334.605 $\pm$ 0.015 	&\cite{PhysRevA.83.042505}	&670\,999\,566.90$\pm$0.89	&-1\,369.88	&\cite{PhysRevA.83.042505} \\ \hline
$^{14,15}$N&$2^1S_0-2^3P_1$& 649\,418\,424.16$\pm$29\,979.2 &\cite{PhysRevA.57.180} & 649\,469\,388.8 $\pm$ 269\,812.8	&	&\cite{PhysRevA.57.180}\\\hline \hline
$^{6,7}$Li&$2^3P_0-2^3S_1$	&3\,474\,773$\pm$55	&\cite{PhysRevA.49.207}&34\,747\,876	&	&\cite{PhysRevA.49.207}\\
 \hline
 \end{tabular}
 \caption{Theoretical and experimental input values for the IS bounds: measured IS $\nu_i^{AA',{\rm exp}}$, theory prediction for a point-like nucleus $\nu_{i,0}^{AA',{\rm th}}$, 
 and the field shift constant $F_i$. See also Ref.~\cite{Pachucki:2017xcg}. For the IS of $2S-12D$ in H/D, see the absolute frequencies in Tab.~\ref{tab:absfreq}. $F_{1S-2S}$ for H/D is an approximate value obtained from the quoted FS in Ref.~\cite{PhysRevA.83.042505}.
 }
 \label{tab:inputIS}
 \end{center}
\end{table*}

\renewcommand{\arraystretch}{1.4}
\begin{table*}[h!]
 \begin{center}
 \begin{tabular}{c|c||c c||c c}
 \hline \hline
 element&transition $i$& $\nu_i^{\rm exp}$\,[kHz] &ref. &$\nu_i^{\rm th}$\,[kHz] &ref\\ \hline\hline
 H	&\multirow{2}{*}{$2S_{1/2}-12D_{3/2,5/2}$}&799\,191\,727\,402.8 $\pm$ 6.7	&\cite{deBeauvoir2000}
 & 799\,191\,727\,409.1 $\pm$ 3.0  &\multirow{2}{*}{\cite{PhysRevLett.95.163003}
 }\\	
 D	&&799\,409\,184\,967.6 $\pm$ 6.5	&\cite{deBeauvoir2000}
 & 799\,409\,184\,973.4 $\pm$ 3.0  &\\	 \hline
   H&\multirow{2}{*}{$(4S_{1/2}-2S_{1/2})-\frac{1}{4}(2S-1S)$}&4797338 $\pm$ 10 	&\multirow{4}{*}{\cite{weitz1995precision}
   } &4\,797\,329 $\pm$ 5 	&\multirow{4}{*}{\cite{weitz1995precision}
   }
   \\
  D&	&4\,801\,693 $\pm$ 20	&	&4\,801\,692 $\pm$ 5	&\\ 
  H&\multirow{2}{*}{$(4S_{1/2}-2D_{5/2})-\frac{1}{4}(2S-1S)$}&6\,490\,144 $\pm$24 	& &6\,490\,128 $\pm$ 5 	&\\
  D&	&6\,494\,841 $\pm$ 41	&	&6\,494\,816 $\pm$ 5	&\\ \hline
  \multirow{4}{*}{$^4$He}
  &$2^1S_0-2^3S_1$&192\,510\,702\,145.6 $\pm$ 1.8	&\multirow{4}{*}{\cite{Pachucki:2017xcg}
  }	&192\,510\,703\,400 $\pm$ 800	&\multirow{4}{*}{\cite{Pachucki:2017xcg}
  } \\
  &$2^3P_0-3^3D_1$&510\,059\,755\,352 $\pm$ 28 	&	&510\,059\,754\,000 $\pm$ 700	& \\
  &$2^3S_1-3^3D_1$&786\,823\,850\,002 $\pm$ 56	&	&786\,823\,848\,400 $\pm$ 1\,300	& \\
  &$2^1S_0-2^1P_1$&145\,622\,892\,886 $\pm$ 183	&	&145\,622\,891\,500 $\pm$ 2\,300	& \\ 
  Ps&$1^3S_1 - 2^3 S_1$&1\,233\,607\,216\,400 $\pm$ 3200	&\cite{Fee:1993zz}	&1\,233\,607\,222\,180 $\pm$ 580	&\cite{Czarnecki:1999mw}
  \\ \hline \hline
 \end{tabular}
 \caption{Measurements and predictions of absolute transition frequencies in H, D, He and positronium (Ps). The H and D values are used for the H/D IS in Sect.~\ref{sec:ISHD}. 
 The middle part of the table summarizes the experimental and theoretical frequencies of the Lamb shift $(4L-2S_{1/2})-\frac{1}{4}(2S-1S)$ for $L=S, D$.
 The lower part of the table is used for constraining $y_e$ in Fig.~\ref{fig:ye}: the transitions in $^4$He with an agreement of better than $2\sigma$ between theory and experiment, as well as a transition in Ps.
 }
 \label{tab:absfreq}
 \end{center}
 \end{table*}

\section{Helium wavefunction} 
\label{app:HeWF}

In this appendix we specify the approximate wavefunctions we use in the helium calculations.
It is conventional to label the states of helium as 
\beq
	n^{2S+1}L_J
\eeq
corresponding to the following electronic configuration $(1s)(nl)$. 
$L$ is the total orbital momentum, $S$ is the spin and $J \le L+S$ is the total angular momentum. 
Since one electron is always in the $(1s)$ orbital $L=l$, and $S=0,1$ corresponding to the singlet and triplet states, respectively. 

For two-electron systems in the non-relativistic limit, the spin and spatial parts of the wavefunction are factorized. 
The spin singlet\,($S=0$) state is 
\beq
	|S=0,m_S=0\rangle = \frac{|\uparrow\downarrow\rangle - |\downarrow\uparrow\rangle}{\sqrt{2}}\,,  
\eeq
while the spin triplet\,($S=1$) is with components
\begin{align}
&	|S=1,m_S=1\rangle = |\uparrow\uparrow\rangle\,,\nonumber \\ 
&	|S=1, m_S=0\rangle = \frac{|\uparrow\downarrow\rangle + |\downarrow\uparrow\rangle}{\sqrt{2}}\,,\nonumber \\
&	|S=1, m_S=-1\rangle = |\downarrow\downarrow\rangle\,.  
\end{align}
Using an antisymmetrized combination of hydrogenic orbitals, the spatial part of the wavefunction takes the form
\begin{align}
	\label{eq:psinlm}
&	\langle  \vec{r}_1,\vec{r}_2|\psi_{nlm}^S\rangle 
=	 \frac{1}{2\sqrt{\pi}}\frac{1}{\sqrt{2(1+N_{nl})}} \times  \nonumber \\
&	  \left[F_{nl}(r_1,r_2) Y_{lm}(\Omega_2)+(-1)^S F_{nl}(r_2,r_1)Y_{lm}(\Omega_1)\right]\,,
\end{align}
where the $1/(2\sqrt{\pi})$ prefactor is the $Y_{00}$ spherical harmonic from the $1s$ electron, and $N_{nl}$ ensures that the radial part of the wavefunction is canonically normalized. We write the $F_{nl}$'s as products of non-relativistic hydrogen radial wavefunctions as
\beq
	F_{nl}(r_1,r_2) = R_{10}(r_1,Z_i) R_{nl}(r_2,Z_a)\,,
\eeq
where (in units of $a_0=1$)
\begin{align}
	R_{nl}(r,Z)
\!=\!	\left(\frac{2Z}{n}\right)^{3/2}  \!\!\!\! \sqrt{\frac{(n-l-1)!}{2n(n+l)!}} 
	e^{-\rho/2}\rho^lL_{n-l-1}^{2l+1}(\rho)\,,
\end{align}
where $\rho=2rZ/n$, $L_k^{\alpha}(x)$ are the generalized Laguerre polynomials of degree $k$, and $Z_i$ and $Z_a$ are the effective nuclear charges for the core $(1s)$ and valence $(nl)$ electrons, respectively. An important point is that $Z_{i,a}\neq Z=2$ because of screening effects; they depend on the electronic configuration considered, see Table~\ref{TableZ}. The $R_{nl}$'s form an orthonormal basis for a fixed $Z$. However, because the two electrons effectively feel a different nuclear charge ($Z_i\neq Z_a$), there is an overall normalization constant for $S$ waves because of the cross term in the square of Eq.~\eqref{eq:psinlm}
\beq
	N_{nl}=(-1)^S\delta_{l,0}\left[\int dr r^2R_{10}(r,Z_i)R_{nl}(r,Z_a)\right]^2\,,
\eeq 
which vanishes for $Z_i=Z_a$ and $n\geq2$ by orthogonality of $R_{nl}(r,Z)$. 

\renewcommand{\arraystretch}{1.4}
\begin{table}[!t]
\center
\begin{tabular}{c |c c}
\hline\hline
state & $Z_i$ & $Z_a$ \\
\hline
\hline
$(2)^1S$ & $2.08$ & $1.21$ \\
$(2)^3S$ & $2.01$ & $1.53$ \\
$(2)^3P$ & $2.00$ & $0.97$ \\
$(2)^3P$ & $1.99$ & $1.09$ \\
$n\geq3$ & $2$ & $1$ \\
\hline
\hline
\end{tabular}
\caption{Effective nuclear charges for the excited states of helium under consideration. $Z_i$ is the charge of the core (1s) electron and $Z_a$ is the charge of the excited electron. These are obtained by variational methods using the non-relativistic hydrogen wavefunctions as trial functions. When the electron is excited to a $n=3$ or higher orbital, the screening of the core electron is found nearly perfect.}
\label{TableZ}
\end{table}
The total wavefunction for a fixed $J$ and (its projection) $m_J=-J\dots J$ are then constructed from $L\times S$ combination of angular momentum using the Clebsch-Gordan coefficients $C_{L,m,S,m_S}^{J,m_J}$ as
\begin{align}
	\label{eq:PsiJmJ}
	|n^{2S+1}L_{J,m_J}\rangle 
= 	\sum_{m=-L}^{L} \sum_{m_S=-S}^S  C_{L,m,S,m_S}^{J,m_J} |\psi_{nLm}\rangle|S,m_S\rangle.
\end{align} 

We use the above wavefunctions for all helium states with the exceptions of the $2^1S$, $2^3S$ and $2^3P$ states where we use non-relativistic wavefunctions based on Hylleraas functions taken from Refs.~\cite{1948ApJ...108..354H,PhysRev.116.914} in order to better describe the repulsion between the two electrons. This turns out to be of particular importance for the $2^1S$ state. Indeed, for spin-singlet states, the spatial part of the wavefunction is symmetric under the exchange of the two-electrons so that the wavefunction in Eq.~\eqref{eq:psinlm} may over-estimate the electronic density is in the region where the electrons are close to each other, $r_1\sim r_2$. Hylleraas functions then provide a more accurate description of the electron repulsion effect by introducing an explicit dependence on the inter-electronic distance $r_{12}$ in the wavefunction. The spatial part of wavefunction is then taken to be of the form
\begin{align}\label{psiHyl}
&\langle \vec{r}_1,\vec{r}_2|\psi_{nlm}^S\rangle = \frac{1}{\sqrt{N}} \left[F_{nl}(r_1,r_2,r_{12})Y_{lm}(\Omega_2)+\right.\nonumber\\
&\left.+(-1)^S F_{nl}(r_2,r_1,r_{12})Y_{lm}(\Omega_1)\right]
\end{align}
where $N$ is a normalization constant and $F_{nl}$ now depends on $r_{12}$ and is expanded on Hylleraas functions as
\beq\label{FHyl}
F_{nl}(r_1,r_2,r_{12}) &=& [\kappa (s+t)]^l e^{-\frac{\kappa}{2}\left(s-\sigma t\right)}\nonumber\\
&&\times\sum_{i=1}^k c_i \phi_i(\kappa s,-\kappa t,\kappa u)\,,
\eeq
with $s\equiv r_1+r_2$, $t\equiv r_2-r_1$, $u\equiv r_{12}$ and $\phi_i(s,t,u)= s^{p_i}t^{q_i}u^{r_i}$. It is convenient to reorganize the Hylleraas terms according to their powers of $r_{12}$. We then write the radial function in Eq.~\eqref{FHyl} as
\beq
F_{nl}(r_1,r_2,r_{12}) =  \sum_{i=0}^k f_i(r_1,r_2) r_{12}^i\,.
\eeq

\section{Overlap integrals for helium} 
\label{app:XYHe}

In this appendix we give the analytical expressions for the overlap integrals for the case of helium, \textit{i.e.\,} the electronic NP coefficients $X_i$ and $Y_i$.

\subsection{Electron-nucleus interactions}

Let us consider the potential of Eq.~\eqref{eq:Ven} between the nucleus and its bound electron with the above helium wavefunctions. In first-order perturbation theory we find
\newpage
\begin{align}
	\label{eq:Xen}
	X_i  \equiv & \hat{X}_a - \hat{X}_b \nonumber\\		
= &\frac{1}{4\pi}\int \left[\prod_{k=1}^{n_e}d^3r_i\right] \,  \left[\sum_{i=1}^{n_e}\frac{e^{-m_\phi r_i}}{r_i}\right]  \nonumber\\
&	\times  \left[ | \Psi_a(\vec{r}_1,\dots,\vec{r}_{n_e}) |^2 - | \Psi_b(\vec{r}_1,\dots,\vec{r}_{n_e}) |^2 \right] \, ,
\end{align}
where $n_e$ is the number of bound electrons and $|\Psi|^2$ is the electron wavefunction density.
Using hydrogenic wavefunctions in Eq.~\eqref{eq:psinlm} the contributions from each state is 
\begin{widetext}
\begin{align}
	\hat{X}_{n^{2S+1}L_{J,m_J}} 
=	&-\frac{1}{4\pi(1+N_{nL})}\left[  \int dr re^{-m_\phi r}\left[R_{10}(r,Z_i)^2+R_{nL}(r,Z_a)^2\right]\right.\nonumber\\
	&+2(-1)^S\delta_{L,0}\int dr re^{-m_\phi r} R_{10}(r,Z_i)R_{nL}(r,Z_a)
	\left.\times \int dr r^2 R_{10}(r,Z_i)R_{nL}(r,Z_a)\right] \, .
\end{align}
\end{widetext}

For the case of Hylleraas wavefunctions in Eq.~\eqref{psiHyl}, we need the following expansion of 
$r_{12}$ raised to the power $k$ on spherical harmonics 
\beq
r_{12}^k = 4\pi \sum_{l=0}^\infty H^{(k)}_l(r_1,r_2) \sum_{m=-l}^l Y_{lm}(\Omega_1)Y_{lm}^*(\Omega_2),
\eeq 
where the coeffcients can be written in closed form in terms of hypergeometric functions~\cite{sack1964generalization}
\begin{align}
&H_l^{(n)}(r_1,r_2) = \frac{(-n/2)_l}{(1/2)_l}\frac{r_>^n}{2l+1}\left(\frac{r_<}{r_>}\right)^l\nonumber\\
&\times F\left[l-n/2,-(n+1)/2;l+3/2;\frac{r_<^2}{r_>^2}\right]\,,
\end{align}
with $F(\alpha,\beta;\gamma;x)$ denoting the Gauss hypergeometric function and $(\xi)_s\equiv \Gamma(\xi+s)/\Gamma(\xi)$. 
We then find
\begin{align}
\hat{X}_{n^{2S+1}L_{J,m_J}} =
&-\frac{1}{4\pi N}\int r_1^2r_2^2dr_1dr_2\left(\frac{e^{-m_\phi r_1}}{r_1}+\frac{e^{-m_\phi r_2}}{r_2} \right)\nonumber\\
&\times \int d\Omega_1d\Omega_2 |\psi_{nlm}^S|^2\,,
\end{align}
where the square of the spatial wavefunction integrated over the angular variables is
\begin{widetext}
\begin{align}
\int d\Omega_1d\Omega_2 |\psi_{nlm}^S|^2 =&f_0(r_1,r_2)^2+\sum_{i+j\geq1} H^{(i+j)}_0(r_1,r_2)f_i(r_1,r_2)f_j(r_1,r_2)+[r_1\leftrightarrow r_2]\nonumber\\
&+2(-1)^S \left[f_0(r_1,r_2)f_0(r_2,r_1)\delta_{l0}+\sum_{i+j\geq1} H_l^{(i+j)}(r_1,r_2)f_i(r_1,r_2)f_j(r_2,r_1)\right]\,.
\end{align}
\end{widetext}

\subsection{Electron-electron interactions}
Consider the NP potential between the bounded electrons, $V_{e}(r_{12})$ see Eq.~\eqref{eq:Ven}.
It is useful to expand the Yukawa potential over spherical harmonics as, see for example~\cite{0953-4075-45-23-235003},  
\begin{align}
	\label{eq:Ylmexpansion}
	\frac{e^{-mr_{12}}}{ r_{12}}
	\!=\! 4\pi\sum_{l=0}^\infty G_l(r_1,r_2,m) \!\!\! \sum_{m=-l}^l \!\!\! Y_{lm}(\Omega_1)Y^*_{lm}(\Omega_2)\,,
\end{align}
where the coefficients are 
\begin{align}
	G_l(r_1,r_2,m)=  \frac{I_{l+1/2}(mr_<)}{\sqrt{r_<}}\frac{K_{l+1/2}(mr_>)}{\sqrt{r_>}}\,,
\end{align}
with $I$ and $K$ the modified Bessel functions of the first and second kind respectively and $r_>$ ($r_<$) is the greater (lesser) of $r_1$ and $r_2$. For Hylleraas wavefunctions which involve additional powers of $r_{12}$ it will be convenient to use Eq.~\eqref{eq:Ylmexpansion} as a ``generating functional" in order to derive the expansion of any $r_{12}^{k-1} e^{-m r_{12}}$ functions (for $k\geq1$) by differentiating $k-$times the coefficients $G_l(r_1,r_2,m)$. 

The first-order perturbation theory result is
\begin{align}
	\label{eq:Ye}
	Y_i \equiv& \hat{Y}_a - \hat{Y}_b \nonumber\\
=& 	\frac{1}{4\pi}\int \left[\prod_{k=1}^{n_e}d^3r_i\right] \,  \left[\sum_{i>j}^{n_e} \frac{e^{-m_\phi r_{ij}}}{r_{ij}}\right] \nonumber\\
&	\times  \left[ | \Psi_a(\vec{r_1},\dots,\vec{r_{n_e}}) |^2 - | \Psi_b(\vec{r_1},\dots,\vec{r_{n_e}}) |^2 \right] \, .
\end{align}

Using the expansion of Eq.~\eqref{eq:Ylmexpansion} and the hydrogenic wavefunctions from Eq.~\eqref{eq:psinlm} we find
\begin{widetext}
\begin{align}
	\label{eq:eeshift}
	\hat{Y}_{n^{2S+1}L_{J,m_J}} 
=	-\frac{1}{4\pi(1+N_{nL})}\int dr_1dr_2 r_1^2r_2^2
	\big[&G_0(r_1,r_2) R_{10}(r_1,Z_i)^2R_{nL}(r_2,Z_a)^2\nonumber\\
	&+(-1)^SG_L(r_1,r_2)R_{10}(r_1,Z_i)R_{10}(r_2,Z_i)R_{nL}(r_1,Z_a)R_{nL}(r_2,Z_a)\big]\,.
\end{align}
\end{widetext}
where only $G_{0,L}$ coefficients in the Yukawa expansion of Eq.~\eqref{eq:Ylmexpansion} are needed. Note that the integrand above no longer depends on $m'$ and $m_S$ hence the sum over Clebsch-Gordan coeffecients squared gives $\sum_{m',m_S}[C_{L,m',S,m_S}^{J,m_J}]^2=1$ by orthonormality. Finally, note that the shift in Eq.~\eqref{eq:eeshift} is indepedent of $J$ and $m_J$, which is expected since the potential in Eq.~\eqref{eq:Ven} is invariant under rotations. We also used the fact the $G_l(r_1,r_2)=G_l(r_2,r_1)$ to simplify the expression.

For the case of Hylleraas wavefunctions in Eq.~\eqref{psiHyl}, we find 
\begin{align}
\hat{Y}_{n^{2S+1}L_{J,m_J}} 
=& -\frac{1}{4\pi N}\int r_1^2r_2^2dr_1dr_2 \nonumber\\
&\times \int d\Omega_1d\Omega_2 \frac{e^{-m r_{12}}}{r_{12}} |\psi_{nlm}^S|^2\,,
\end{align}
where the angular integral simplifies to
\begin{widetext}
\begin{align}
\int d\Omega_1d\Omega_2  \frac{e^{-m r_{12}}}{r_{12}} |\psi_{nlm}^S|^2 =&\sum_{ij} G^{(i+j)}_0(r_1,r_2)\left[f_i(r_1,r_2)f_j(r_1,r_2)+f_i(r_2,r_1)f_j(r_2,r_1)\right]\nonumber\\
&+2(-1)^S \sum_{ij} G_l^{(i+j)}(r_1,r_2)f_i(r_1,r_2)f_j(r_2,r_1)\,,
\end{align}
\end{widetext}
where $^{(k)}$ indicates the $k$th differentiation with respect to $m$,
\beq
G_l^{(k)}(r_1,r_2,m)\equiv (-1)^k \frac{\partial^{k}G_l(r_1,r_2,m)}{\partial m^k}\,. 
\eeq

\begin{figure}[ht!]
 \begin{center}
  \includegraphics[width=0.9\columnwidth]{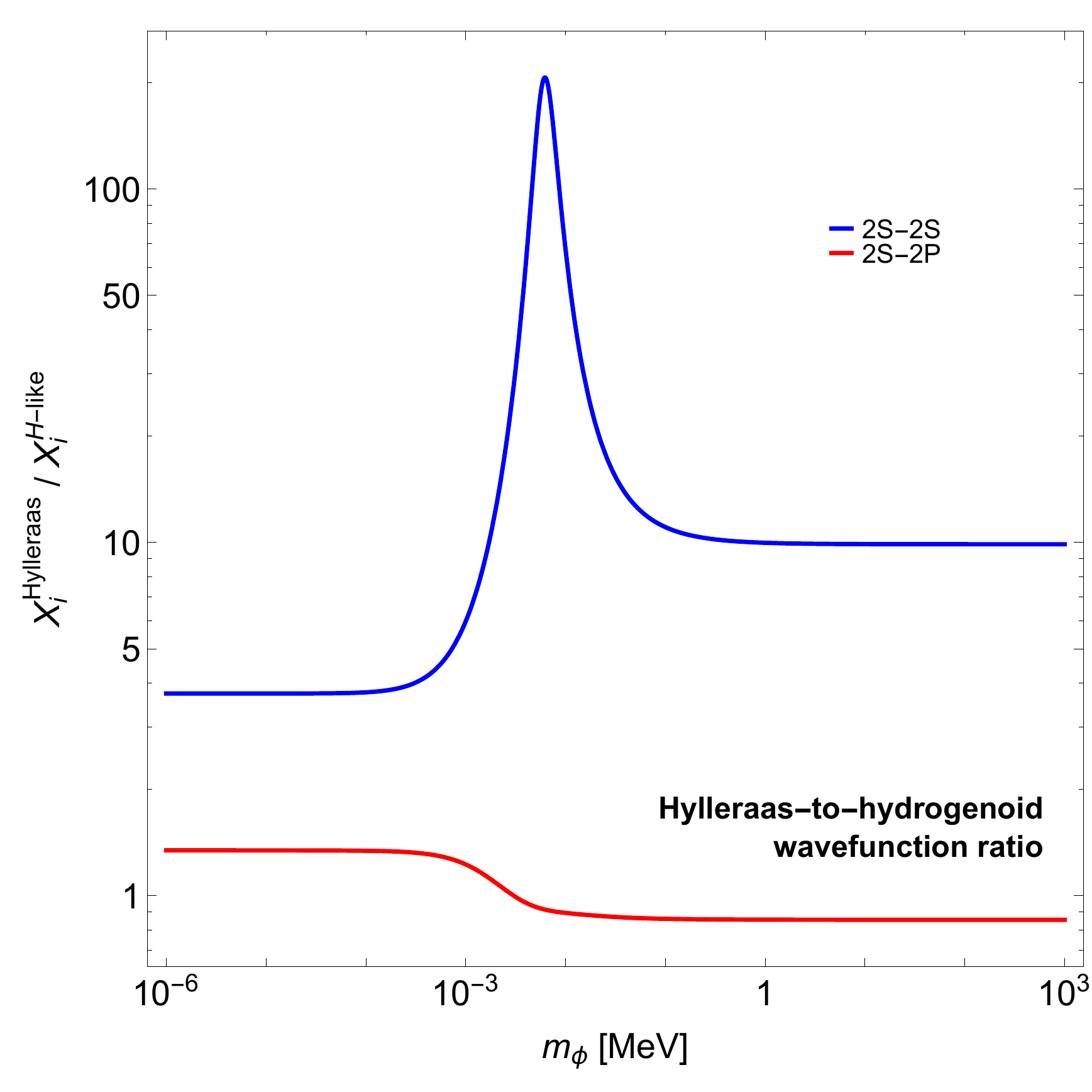}
  \includegraphics[width=0.9\columnwidth]{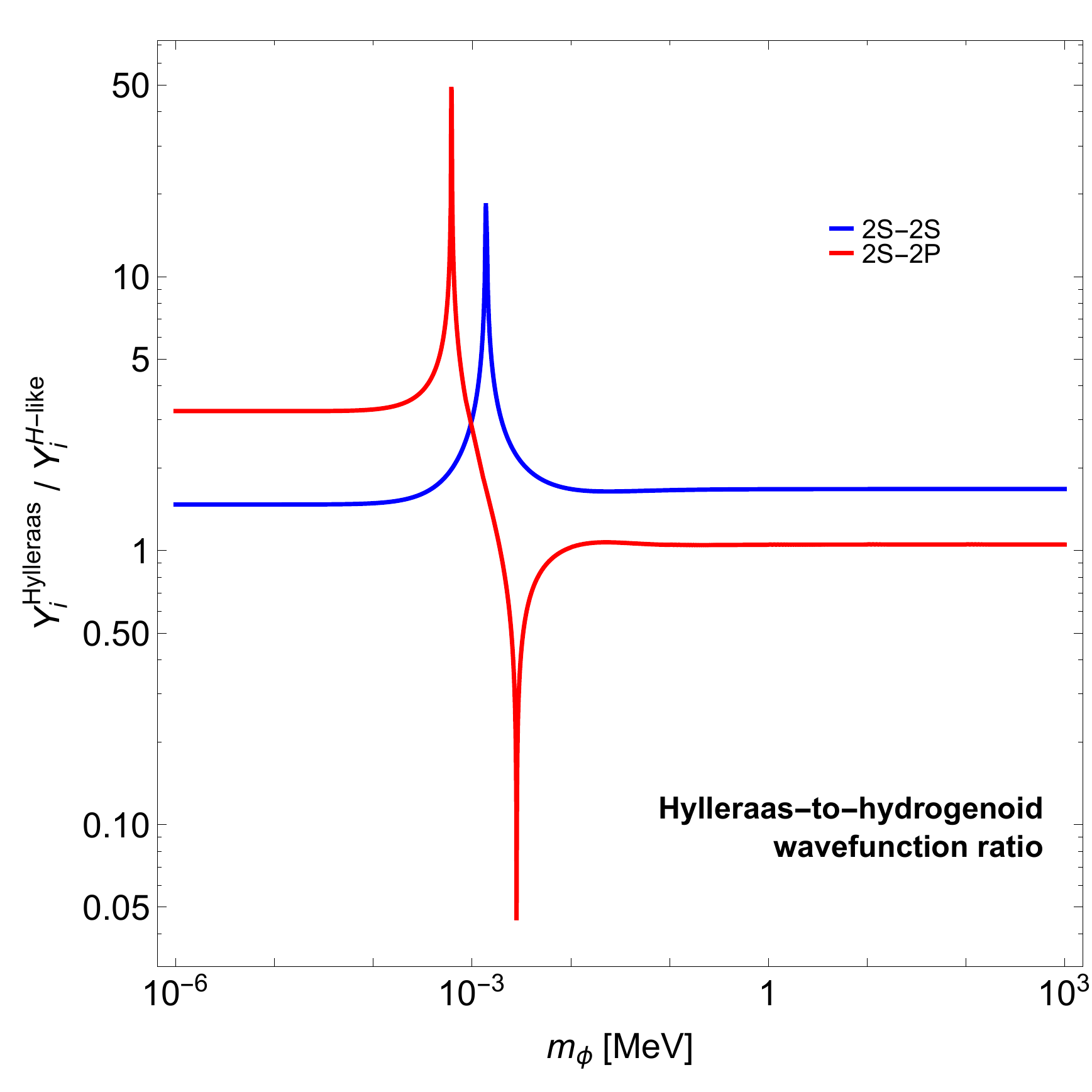}
  \caption{Comparison of the electronic NP coefficients $X_i,~\,Y_i$ based on hydrogen-like and Hylleraas wavefunctions depending on the mediator mass $m_\phi$ for the transitions $2S-2S$ and $2P-2S$ in helium. }
  \label{fig:HHylleraas}
 \end{center}
\end{figure}

In Fig.~\ref{fig:HHylleraas} we evaluate the impact of the Hylleraas wavefunctions on the electronic NP constants by calculating the ratios to the respective quantity based on hydrogen-like wavefunctions,
$X_i^{\rm Hylleraas}/X_i^{\rm H-like}$, 
$Y_i^{\rm Hylleraas}/Y_i^{\rm H-like}$.

\bibliographystyle{utphys}
\bibliography{lightatoms-bib}

\end{document}